\documentclass[12pt]{iopart}
\usepackage{graphicx}
\graphicspath{%
    {converted_graphics/}
    {/}
}
\begin{document}

\title{Cooperative sequential adsorption models on a Cayley tree: analytical results and applications}

\author{D. A. Mazilu, I. Mazilu, A. M. Seredinski, V. O. Kim, B. M. Simpson, W. E. Banks}

\address{Department of Physics and Engineering, Washington and Lee University, Lexington, VA, 24450}
\ead{mazilui@wlu.edu}

\begin{abstract}
We present a class of cooperative sequential adsorption models on a Cayley tree with constant and variable attachment rates and their possible applications for ionic self-assembly of thin films and drug encapsulation of nanoparticles.
Using the empty interval method, and generalizing results known from reaction-diffusion processes on Cayley trees, we calculate a variety of quantities such as time-dependent surface coverage and time-dependent probabilities of certain particle configurations.  
\end{abstract}

\maketitle

\section{Introduction}

Nanotechnology research is a highly interdisciplinary area of study that relies on fundamental understandings of physics, chemistry, biology and materials science. Layer-by-Layer self-assembly of nanoparticles is an  innovative,  widely used technique in nanotechnology studies with applications in microelectronics, optical coatings and biology \cite{heflin}. In this context, non-equilibrium statistical physics methods and models can shed some light on the cooperative behavior of multi-particle systems. 

One-dimensional sequential adsorption models have been studied thoroughly in different physics contexts \cite{evans}, \cite{privman}. The two basic one-dimensional models are known in literature as random sequential adsorption (RSA), when the adsorption sites are chosen randomly, and cooperative sequential adsorption (CSA), when the adsorption mechanism is influenced by the local environment. Despite numerous studies of one-dimensional models, adsorption in two dimensions is not as well understood. There are quite a few computational papers \cite{evans} on the matter, but very few analytical solutions exist for the general two-dimensional case.
 
The adsorption of particles is exactly solvable in higher dimensions only for a finite tree-like lattice called a Cayley tree. A Cayley tree is a connected, cycle-free graph with each node connected to $z$ neighbors, where $z$ is called the coordination number (Fig. 1). For $z = 2$, the tree reduces to the one-dimensional case, i.e., a line. It has been proven that a Cayley tree structure with $z=4$ well approximates a regular two-dimensional lattice for certain cases, such as monomer and dimer deposition. For example, the Cayley tree predicts an 88.9\% dimer coverage while the regular lattice predicts a 90.8\% dimer coverage (result obtained from computer simulations \cite{evans}). Recently, analytical results were reported for the random sequential process \cite{cadilhe} and reaction-diffusion processes on Cayley trees and Bethe lattices \cite{matin},\cite{ali},\cite{gouet}.

In this paper, we study a cooperative sequential model of monomer  and dimer deposition with both constant and variable attachment rates on a Cayley tree. Two experimental topics motivate our paper. One is the understanding of the self-assembly  mechanism of charged nanoparticles on a glass substrate \cite{iler}. Known in literature as ionic self-assembled monolayers (ISAM), this technique has been used successfully in the making of antireflective coatings \cite{ritter}. The properties of these coatings depend on the surface coverage of the substrate. During the manufacturing process, it is highly desirable to know the analytical relationship between the index of refraction and the particle density of the surface. We study the ionic self-assembly deposition from the point of view of stochastic cooperative sequential adsorption models on a Cayley tree lattice. For the purpose of building a theoretical model for this process, we emphasize some of its essential elements: i) The deposition process is stochastic, so a random sequential adsorption model is justified; ii) The nanoparticles deposited are electrically charged, so a cooperative sequential adsorption model with rates dependent on the nearest neighbor occupation is appropriate; iii) The surface on which particles are being deposited is covered with a polyelectrolyte. We consider a Cayley tree as an approximation for this surface. This choice is particularly appropriate when the polymer used is poly(amidoamine) (PAMAM), which has by design a tree-like structure. For different polyelectrolytes, such as poly(allylamine hydrochloride) (PAH) or poly(diallyldimethylammonium chloride) (PDDA), this approximation works well enough for special cases that will be addressed in section 3.

The other experimental interest involves the versatile properties of  synthetic polymers called 'dendrimers", with potential use as a novel drug delivery mechanism via drug encapsulation \cite{dendri1},\cite{dendri2}. Dendrimers are perfect physical examples of Cayley tree structures. They are highly branched, spherical polymers that consist of hydrocarbon chains with variable functional groups attached to a central core molecule. Due to the precise control that can be exerted over their size, molecular architecture, and chemical properties, dendrimers have great potential in the pharmaceutical industry as effective carriers for drug molecules. These new synthetic polymers are able to carry both targeting molecules and drug molecules to cancerous tumors, minimizing the negative side effects of medications on healthy cells. In order to describe mathematically the encapsulation process of drug molecules, we propose a stochastic deposition model with variable attachment rates.
 
Our paper is structured as follows. In section 2 we define our general model for both constant and variable rates. In section 3 we discuss in more depth the ionic self-assembly mechanism and calculate the quantities of interest,
such as the coverage of the surface in the steady state, the probability of having clusters of connected empty sites of a certain size, and the probability of having a certain shell particle distribution. In section 4 we address the drug encapsulation model and present a set of analytical results for the case of variable attachment rates. We conclude in section 5 with a summary of our results and some open questions.

\section{Model definition}

Random sequential adsorption of a mixture of monomers and dimers on a Bethe lattice was solved by Cadilhe and Privman \cite{cadilhe1}, \cite{cadilhe}. Using the empty interval method, they found exact analytical expressions for the time evolution of clusters of connected empty sites. We address the question of adsorption of monomers and dimers with cooperative effects due to different neighbor occupation. We consider the following general processes defined on a Cayley tree:

\begin{eqnarray*}
\bullet \circ &\rightarrow& \bullet \bullet \mbox{ with rate $r_1$}\\
\circ \circ &\rightarrow& \bullet \circ \mbox{ with rate $r_2$}\\
\circ \circ &\rightarrow& \bullet \bullet \mbox{ with rate $r_3$}\\
\end{eqnarray*}
where ''$\bullet$" are filled sites and ''$\circ$" are empty sites. The same processes, with constant, positive rates, were considered in the context of diffusion-reaction problems by Matin \it et al.\rm  \cite{matin} and Alimohammadi \it et al. \rm  \cite{ali}. 

In the following sections, we consider two distinct cases for the deposition rates $r_1$, $r_2$ and $r_3$: i) constant rates, to model the ionic self-assembly process; ii) variable rates, to model the drug encapsulation mechanism using dendrimers. 

For both cases, some general information regarding Cayley trees is needed. According to the mathematical definition, a Cayley tree of order $z$ is defined in the following way \cite{ostilli}. Given a root vertex $0$, we link this vertex to $z$ new vertices by means of $z$ edges.  Shell $\ell=1$ is made of the first set of $z$ new vertices. Each vertex on shell $\ell=1$ is linked to $z-1$ new vertices, and so on. Fig. 1 shows the first 3 shells for a Cayley tree of coordination number $z=4$. Cayley trees are finite trees with boundaries defined by the last shell; in the case presented in Fig. 1 this is shell $L=3$. The number of vertices on the $\ell^{th}$ shell is $n_{\ell}=z(z-1)^{\ell-1}$, and the total number of vertices of the Cayley tree with $L$ shells is $N=z((z-1)^{L}-1)/(z-2)$.

We calculate the following quantities: 
\begin{itemize}
  \item $E_n(t)$, the probability of finding a cluster of $n$-connected empty sites inside the tree, \it regardless \rm of the occupation of the rest of the nodes.
  \item $P_L(m;t)$, the probability of having $m$ particles on the last shell $L$, with all other shells empty. 
  \item $P_{\ell}(m;t)$, the probability of having $m$ particles on the shell $\ell<L$, with all other shells empty. 
  \item $P(m_1;m_2;..m_{\ell};m_L;t)$, the probability of having $\{m_1;m_2;..m_{\ell}\}$ particles distributed on interior shells and $m_{L}$ particles on the last shell, provided that there is one shell empty between filled levels.
\end{itemize}

\section{Simple model of ionic self-assembly: cooperative sequential adsorption with constant rates}

Ionic self-assembly is a relatively new technique \cite{isam1}, \cite{isam2} that allows detailed structural control of materials at the molecular level, combined with ease of manufacturing and low cost. The ISAM process allows one to deposit alternating layers of cations and anions by dipping the substrate in aqueous solutions of the appropriate ions, as illustrated in Fig. 2. Because it is a dipping process, any exposed surface is homogeneously coated, allowing highly uniform, conformal coatings on irregular shapes. The basic building block for the film is a cation/anion bilayer, which may consist of either two polyelectrolytes (a polycation and a polyanion), or a polyelectrolyte and a nanoparticle, or two different nanoparticles. The thickness of a bilayer is a function of the diameter of the nanoparticle and the packing of the particles from layer to layer. Fig. 3 shows a $SiO_2/PDDA$ bilayer made with spherical silica particles with nominal sizes of approximately 45 nm, self-assembled on a glass slide. Some dimers are visible as well.  

We model the ISAM mechanism as a cooperative sequential adsorption of nanoparticles with constant deposition rates $r_1$, $r_2$ and $r_3$ on a Cayley tree. We performed Monte Carlo simulations for both a Cayley tree of coordination number $z=4$ and a fully two-dimensional lattice and monitored the filling process of both lattices. The simulation results show that for similar values of $r_1$ and $r_2$, there is a very good match between a Cayley tree and a fully two-dimensional lattice. The results start to differ once $r_1$ and $r_2$ are significantly different. The attachment rates can be picked to fit the experimental data. A $pH=9$, for example, translates into a higher attachment rate for monomers than $pH=8$. This model was discussed in the context of reaction-diffusion reactions on a Cayley tree by Matin \it et al. \rm \cite{matin} and Alimohammadi \it et al. \rm \cite{ali}. 

\subsection{Calculation of probability distribution of clusters of n-connected empty sites using the empty interval method}

For the ionic self-assembly process, one of the main concerns is the time dependent surface coverage, and its properties in the steady state. Let $P(\bigcirc_n)=E_{n}$ be the probability of finding a cluster of $n$ connected empty nodes, \it regardless \rm of the state of the bordering sites. For clarity, in Fig. 4(a) we present a cluster of four connected empty sites, $E_4$. From now on, we assume that an "$n$-cluster" refers to a cluster of $n$ connected sites, unless specified otherwise. Let $P(\bullet-\bigcirc_n)$ be the probability to have an $n$-cluster that is empty connected to an occupied node via one single link. In this case, we have the following conservation of probability equation \cite{ben avraham}:
\begin{equation}
P(\bullet-\bigcirc_n)+P(\circ-\bigcirc_n)=P(\bigcirc_n)\\
\end{equation}
 which leads to 
\begin{equation}
P(\bullet-\bigcirc_n)=E_n-E_{n+1}
\end{equation}

The equation that governs the probability of having an $n$-cluster of empty sites is:
\begin{eqnarray}
\frac{d E_n}{dt} =-(n(z-2)+2)(r_1P(\bullet-\bigcirc_n)+(r_2+r_3)E_{n+1})-(2r_2+r_3)(n-1)E_n \nonumber
\\
\end{eqnarray}
which can be simplified to:

\begin{eqnarray}
\frac{d E_n}{dt} =-[r_1(n(z-2)+2)+(2r_2+r_3)(n-1)]E_{n}-(r_2+r_3-r_1)(n(z-2)+2)E_{n+1}\nonumber
\\
\end{eqnarray}
For the special case of $r_1=r_2+r_3$, and a completely empty lattice at the initial time, the solution for $E_n$ is:

\begin{equation}
E_{{n}} \left( t \right) ={{\rm e}^{- \left( n \left( r_{{1}} \left( z
-2 \right) +2\,r_{{2}}+r_{{3}} \right) +2\,r_{{1}}-2\,r_{{2}}-r_{{3}}
 \right) t}}
\end{equation}

For the general case of $r_1\neq r_2+r_3$, using the standard method presented in \cite{redner}, we obtain the same results as  Matin \it et al. \rm \cite{matin}. We use the following exponential ansatz:

\begin{equation}
E_n(t)=E_{1}(t)(\phi(t))^{n-1}
\end{equation}

We substitute this into Eq. 4, and we get a system of coupled differential equations by separating the initial equation in terms linear in $n$ and independent of $n$:

\begin{eqnarray}
\frac{d\phi(t)}{dt}&=&-\beta\phi(t)-\beta\alpha \phi(t)^2\\
\frac{dE_1(t)}{dt}&=&-(zr_1+\frac{z}{z-2}\alpha\beta\phi(t))E_1(t)
\end{eqnarray} 

with the notations:
\begin{eqnarray}
\beta=(z-2)r_1+2r_2+r_3\\
\alpha=\frac{(z-2)(r_2+r_3-r_1)}{(z-2)r_1+2r_2+r_3}
\end{eqnarray}

With the appropriate initial conditions ($E_n(t=0)=1$, empty lattice), we recover the following results from \cite{matin}:

\begin{eqnarray}
E_1(t)&=&e^{-zr_1t}(\frac{1}{1+\alpha(1-e^{-\beta t})})^{\frac{z}{z-2}}\\
E_n(t)&=&e^{-zr_1t-(n-1)\beta t}(\frac{1}{1+\alpha(1-e^{-\beta t})})^{\frac{z}{z-2}+n-1}
\end{eqnarray}

Fig. 4(b) displays the probability of finding a cluster of four empty, connected nodes ($n=4$) for a Cayley tree with $z=3$, and arbitrary deposition rates  $r_1 = 0.2$, $r_2 = 0.4$, $r_3 = 0.1$. The initial lattice is empty, and the probability approaches zero in the steady state.

We can also answer the question of the particle density of the jammed state. When all three attachment rates are non-zero, the final state will be completely full. For the case of $r_1=0$, for example, the final state is

\begin{equation}
\rho=1-E_1=1-  \left( {\frac { \left( r_{{2}}+r_{{3}} \right)  \left( z-2 \right) {
{\rm e}^{- \left( 2\,r_{{2}}+r_{{3}} \right) t}}+r_{{3}}-z \left( r_{{
2}}+r_{{3}} \right) }{-2\,r_{{2}}-r_{{3}}}} \right) ^{-{\frac {z}{z-2}
}}
\end{equation}

In Fig. 5(a) we present the time dependence of the particle density for $z=4$, $r_1=0$, $r_2=1$, and $r_3=0.1$. The final coverage is 76\%. In Fig. 5(b) we have the steady state coverage ($t\rightarrow \infty$) as a function of coordination number $z$ for the same set of parameters. The coverage increases with time, and tends asymptotically to 100\% as $z$ grows larger. 

\subsection{Probability distributions of shell occupation}
For the ionic self-assembly mechanism, it is interesting to know not only the final surface coverage, but also the probability of having certain particle patterns. The results presented below are applicable for the case of a ``snowflake" type polymer, such as PAMAM. We assume below an ideal case of  particle deposition on only one "snowflake" of generation $L$.
Alimohammadi \it et al. \rm \cite{ali} addressed the possible reaction-diffusion processes on a Cayley tree, and showed that the only reactions amenable to exact solutions are the ones listed in our model. Below, we discuss some of their results in the context of particle deposition with constant rates, and we also present a more general solution to the problem. 
For convenience, we remind the reader of the quantities of interest:
\begin{itemize}
\item $P_L(m;t)$, the probability of having $m$ particles on the last shell $L$, with all other shells empty. 
 \item $P_{\ell}(m;t)$, the probability of having $m$ particles on the shell $\ell<L$, with all other shells empty. 
 \item $P(m_1;m_2;..m_{\ell};m_L;t)$, the probability of having $\{m_1;m_2;..m_{\ell}\}$ particles distributed on interior shells and $m_{L}$ particles on the last shell, provided that there is one shell empty between filled levels.
\end{itemize}
The empty shell between filled consecutive shells is necessary in order to ensure a closed set of differential equations \cite{ali}.\\

 \bf a) Internal shells $\ell<L$ \rm
\\

The master equation that governs particle deposition is:

\begin{equation}
\frac{\partial P_{\ell}(m;t)}{\partial t}=(n_{\ell}-m+1)zr_{2}P_{\ell}(m-1;t)-[mzr_1+(2r_2+r_3)(N-mz)]P_{\ell}(m;t)
\end{equation}

with $n_{\ell}$ being the number of sites on the $\ell^{th}$ shell and $N=\sum_{\ell'=1}^{L} n_{\ell'}$ the total number of sites.

\begin{equation}
n_{\ell} = \left\{ \begin{array}{lrc}
 z(z-1)^{\ell-1} & \mbox{for} & \ell\geq 1 \\
1 & \mbox{for} & \ell=0
\end{array}\right.
\end{equation}

The first term on the right hand side of Eq. 14 accounts for the processes that lead to configuration $P_{\ell}(m;t)$. Starting with $m-1$ particles on shell $\ell$, there are $(n_{\ell}-m+1)z$ possibilities for a particle to be deposited on shell $\ell$. The negative terms in the master equation account for the disappearance of configuration $P_{\ell}(m;t)$ via particle depositions on shell $\ell$ or on the other interior shells.

For simplicity, we introduce the following notations:
\begin{eqnarray}
a=r_1-2r_2-r_3\\
b=N(2r_2+r_3)
\end{eqnarray}
With these notations, Eq. 14 can be rearranged as:
\begin{equation}
\frac{\partial P_{\ell}(m;t)}{\partial t}=(n_{\ell}-m+1)zr_{2}P_{\ell}(m-1;t)-(mza+b)P_{\ell}(m;t)
\end{equation}

The method used is recursive, similar to \cite{ali}.  The starting point is the case of $m=0$, when the equation becomes:

\begin{equation}
\frac {dP_{\ell}(0;t)}{dt} =-bP_{\ell} (0;t)
\end{equation} 

We assume an empty initial configuration, so $P_{\ell}(0;0)=1$. The solution is:

\begin{equation}
P_{\ell}(0;t)={{\rm e}^{-bt}}
\end{equation} 
The solution for $P_{\ell}(0;t)$ becomes the "starter" for the equation for $P_{\ell}(1;t)$ and so on, to get the following general solutions:

\begin{eqnarray}
P_{{\ell}}(m;t)&=&{\frac {{r_{{2}}}^{m} \left( -1 \right) ^{m}\Gamma  \left( m-{
\it n_{\ell}} \right) {{\rm e}^{-bt}} \left( 1-{{\rm e}^{-zat}} \right) ^{m}
}{\Gamma  \left( -{\it n_{\ell}} \right) m!\,{a}^{m}}}
  \mbox{    for $a\neq0$}\\
P_{{\ell}}(m;t)&=&{\frac {({r_{{2}}tz})^{m}\left( -1 \right) ^{m}
\Gamma  \left( m-{\it n_{\ell}} \right) {{\rm e}^{-bt}}}{m!\,\Gamma  \left( 
-{\it n_{\ell}} \right) }}
  \mbox{   for $a=0$}
\end{eqnarray}

Fig. 6(a) shows the time dependence of the probability of having four particles in the third shell of five, with all other shells empty, for $z=4$, $r_1=0$, $r_2=1$, and $r_3=0.1$. The probability reaches a maximum of roughly $7.08\times10^{-5}$ at $t=4\times 10^{-3}$ in arbitrary units.  

Larger $r$-values shrink the time scale and condense the curve. Smaller $r$-values elongate the time scale, that is, the lower the rate of deposition, the longer it will take to get to the desired number of particles in the shell. Of the three, $r_2$ and $r_3$ have the greatest influence over the curve, altering both the time scale and the order of magnitude of the probability for the particle distribution in the shell. We can also find analytically $t_{max}$, the peak of the distribution:
\begin{equation}
t_{max}=-\frac{\ln\frac{((z-1)^{L}-1)\frac{b}{N}}{\frac{b}{N}((z-1)^{L}-1)+ma(z-2)}}{za}
\end{equation}

\bf b) Last shell $L$ \rm
\\
The master equation that governs these processes is (set $z=1$ in Eq.14):
\begin{equation}
\frac{\partial P_{L}(m;t)}{\partial t}=(n_{L}-m+1)r_{2}P_{L}(m-1;t)-(ma+b)P_{L}(m;t)
\end{equation}
with the solutions:
\begin{eqnarray}
P_{{L}}(m;t)&=&{\frac {{r_{{2}}}^{m} \left( -1 \right) ^{m}\Gamma  \left( m-{
\it n_{L}} \right) {{\rm e}^{-bt}} \left( 1-{{\rm e}^{-at}} \right) ^{m}
}{\Gamma  \left( -{\it n_{L}} \right) m!\,{a}^{m}}}
 \mbox{  for $a\neq0$}\\
P_{{L}}(m;t)&=&{\frac {({r_{{2}}t})^{m}\left( -1 \right) ^{m}
\Gamma  \left( m-{\it n_{L}} \right) {{\rm e}^{-bt}}}{m!\,\Gamma  \left( 
-{\it n_{L}} \right) }}
 \mbox{ for $a=0$}
\end{eqnarray}

Fig. 6(b) shows the time dependence of the probability of having four particles in the final ($L=5$) shell with all other shells empty for $z=4$, $r_1=0$, $r_2=1$, and $r_3=0.1$. The probability reaches a maximum of roughly $2.01\times10^{-3}$ at $t=3.95\times10^{-3}$. Notice the overall similarity of this curve to the curve for interior shells. The relationships between the $r$-values and the curve are the same as in Fig. 6(a). Note that the order of magnitude for the probability here is $10^{-3}$ or less, compared to $10^{-5}$ for an interior shell. This is the result of the outer shell always having the most possible spaces to occupy, so it is more likely that it will fill to a certain point with the rest of the tree empty than for any interior shell to fill to that same point under the same conditions. As before, we can find an analytical expression for the time when the maximum of the distribution is reached:
\begin{equation}
t_{max}=-\frac{\ln\frac{z((z-1)^{L}-1)\frac{b}{N}}{z((z-1)^{L}-1)\frac{b}{N}+ma(z-2)}}{a}
\end{equation}

\subsection{General solution for $P(m,t)$}

Consider a case of $m=m_1+m_2+m_3+..$ particles being distributed on interior shells with at least one shell empty between them. We want to calculate the probability of this particle distribution as a function of time.

The general form for the equation for one shell occupied with all others empty can be expressed as: 
\
\begin{equation}
 \frac{\partial P(m_{\ell} )}{\partial t} = r_2 z (n_{\ell}- m_{\ell}+1) P(m_{\ell} -1)-(mza+b)P(m_{\ell} ) , \label{Pmell}
 \end{equation}

for any inner shell $\ell$.\\
Using Eq. \ref{Pmell} as an example, we move the $b$ term from the right to the left side of the equal sign, allowing us to rewrite this equation as:
\begin{eqnarray}
 \frac{\partial P(m_{\ell} )}{\partial t}+bP(m_{\ell}) &=& \exp(-b t) \frac{\partial (\exp(bt) P(m_{\ell} ))}{\partial t}\\ \nonumber
 &= &r_{2} z (n_{\ell}- m_{\ell}+1) P(m_{\ell} -1)-m z aP(m_{\ell}) . 
\end{eqnarray}\\
Noting that the time dependence of the totally unoccupied structure is:
\[ P(0) = P_L(0) = P(0,0,\cdots) = \exp(-b t) , \]

and defining the time-dependent functions $Q(m_{\ell}) \equiv P(m_{\ell}) / P(0)$ , 
we can rewrite Eqns. 25 and \ref{Pmell} in terms of the $Q$'s as:

\begin{equation} 
\frac{\partial Q(m_{\ell})}{\partial t} = r_{2}z(n_{\ell}- m_{\ell}+1) Q(m_{\ell} -1)- mza Q(m_{\ell} )  \label{QL} 
\end{equation}

For the case of an outer shell, the equation is:
\begin{equation} 
\frac{\partial Q(m_{L})}{\partial t} = r_{2}(n_{L}- m_{L}+1) Q(m_{L} -1)- ma Q(m_{L} )  \label{QL} 
\end{equation}

Now consider the case of three occupied shells -- two inner shells and the outer shell separated by at least one totally empty shell -- and examine the equation satisfied by the product $Q(m_l, m_k, m_L) \equiv Q(m_l) Q(m_k) Q_L(m_L) $.

\begin{eqnarray}
\frac{\partial Q(m_{\ell}, m_k, m_L) }{\partial t} =\frac{\partial Q(m_{\ell} )}{\partial t}Q(m_k) Q_L(m_L) + \frac{\partial Q(m_k )}{\partial t}Q(m_{\ell}) Q_L(m_L) +\nonumber \\
\frac{\partial Q_L(m_L )}{\partial t}Q(m_{\ell}) Q_L(m_k) \nonumber \\
=  r_{2} z (n_l+1 - m_l) Q(m_l-1, m_k, m_L) + r_{2} z (n_k+1 - m_k)Q(m_l, m_k-1, m_L) \nonumber \\ 
   +  r_{2}(n_L+1 - m_L) Q(m_{\ell}, m_k, m_L-1) - (az(m_{\ell} + m_k) + zm_L) Q(m_{\ell}, m_k, m_L) \nonumber\\
\end{eqnarray}

This is completely equivalent to the master equation written directly for three occupied shells upon making the substitution $Q(m_l, m_k, m_L) = P(m_l)P(m_k)P_L(m_L) / P(0)^{3}$.
This procedure is trivially extended to cover any number of occupied shells, thus generalizing the following result:
\begin{equation}
P(m_1;m_2;..m_{\ell};m_L;t)=\frac{\prod_{i\in\{1,2,..\ell\}} P(m_i;t)}{P(0;0;t)^{L_{total}}}
\end{equation}

In the set $\{1,2,..\ell\}$, $``1"$ is the first occupied shell, $``2"$ is the second occupied shell, etc., with the assumption that there is always at least one empty shell between occupied shells. $L_{total}$ is the total number of occupied levels (including the last external shell).

\section{Nanoparticle deposition with variable attachment rates}

The shell-dependence of the deposition rates is particularly relevant for the drug encapsulation mechanism via dendrimers. There are two methods of drug delivery using dendrimers: attachment of drug molecules to the outer functional groups and the encapsulation of drugs within the cavities created by separate branches. For example, dendrimer surface chains can form covalent bonded complexes with anti-cancer molecules such as cisplatin \cite{dendri1}. The release of these molecules can be controlled by manipulating the rate at which these dendrimer-drug bonds are degraded. The potential load of each dendrimer carrier can be easily varied by adjusting the branch multiplicity of the dendrimer generation. Another mechanism of dendrimers to load molecules is through molecular encapsulation. The branches of the dendrimers form a dendritic box around the encapsulated molecule, which can protect sensitive molecules from unfavorable physiological environments, and vice-versa. 

The encapsulation process is complex, and depends on the type of dendrimer used, the type of drug molecules encapsulated, and other physical or chemical factors. We propose a minimalist model that considers the drug molecules as generic monomers capable of attaching to the available nodes. To capture the fact that it is less likely to have drug molecules present in the outer shells, we define shell-dependent attachment rates. We treat the surface of the dendrimer separately, since all nodes are available for attachment. Fig. 7 presents a sketch of the encapsulation of drug molecules and attachement of drug molecules onto the surface of a dendrimer. Dendrimers can carry different types of nanoparticles (illustrated by the different shapes in particles in the picture).

In order to model the encapsulation mechanism of drug molecules much larger than the dendrimer nodes, we consider here a special case where the deposition of monomers is forbidden if the neighboring connected site is occupied. Dimer deposition is not being considered. The long time behavior of such a model is more interesting because eventually the system will reach a "jammed" state with a certain percentage of empty sites. Because of their size and their electrostatic interactions, drug molecules cannot be in close proximity, and this fact is reflected in our model by obstructing occupation of adjacent connected sites. Also, inside the dendrimer tree, the closer to the core the drug molecules are, the more available space they have, and the more likely they are to be there. For our model, this translates into shell-dependent deposition rates that decrease for shells further from the core.
For the drug encapsulation model, what is physically relevant is the particle distribution on the shells. We use the same analytical method as the one for the constant rates case.

\subsection{Probability distributions of shell occupation}

\bf a) Monomer deposition on the last shell $L$: variable rates \rm
\\

The following master equation describes the monomer deposition on the last shell $L$.
\begin{eqnarray}
\frac{\partial P_L(m;t)}{\partial t}&=&r(L)(n_L-m+1)P_L(m-1;t)-[r(L)( n_L-m)\\
&+&\sum_{l'=1}^{L-1} n_{l'}(r(l')+r(l'-1))]P_L(m;t)\nonumber
\end{eqnarray}
 with $n_l$ being the number of sites on the $\ell^{th}$ shell.

\begin{equation}
n_l = \left\{ \begin{array}{lrc}
 z(z-1)^{l-1} & \mbox{for} & l\geq 1 \\
1 & \mbox{for} & l=0
\end{array}\right.
\end{equation}
The first term on the right hand side accounts for the processes that lead to configuration $P_L(m;t)$. Starting with $m-1$ particles on layer $L$, there are $(n_L-m+1)$ possibilities for a particle to be deposited on shell $L$. The negative term in the master equation accounts for the disappearance of configuration $P_L(m;t)$ via particle depositions on shell $L$ or on the interior shells $l=1...L-1$.

We assume that $r(L)=1$, and for now we consider for the rest of the shells $r(l)=\lambda^{l}$, with $\lambda<1$ as a function of the shell number. The choice of $r(l)$ is consistent with electrostatic screening and with the distribution of drug molecules encapsulated in the dendrimer.

\begin{eqnarray}
\frac{\partial P_L(m;t)}{\partial t}&=&(n_L-m+1)P_L(m-1;t)-[( n_L-m)+\\\nonumber
&+&\sum_{l'=1}^{L-1} n_{l'}(r(l')+r(l'-1))]P_L(m;t)
\end{eqnarray}
The sum on the right hand side can be written as:
\begin{equation}
\sum_{l'=1}^{L-1} n_{l'}(r(l')+r(l'-1))=z(\lambda+1)\sum_{l'=1}^{L-1}(\lambda(z-1))^{l'-1}\\
\end{equation}

If we work under the assumption that $\lambda<\frac{1}{z-1}$, then the sum in Eq. 37 is the partial sum of a geometric series:

\begin{eqnarray}
z(\lambda+1)\sum_{l'=1}^{L-1}(\lambda(z-1))^{l'-1}=z(\lambda+1)\frac{1-(\lambda (z-1))^{L-1}}{1-\lambda(z-1)}\nonumber
\\
\end{eqnarray}

Eq. 36 can be written in a more compact form,

\begin{equation}
\frac{\partial P_L(m;t)}{\partial t}=(n_L-m+1)P_L(m-1;t)-[( n_L-m)+\frac{(\lambda+1)(z-n_L\lambda ^{L-1})}{1-\lambda(z-1)}]P_L(m;t)
\end{equation}
The general solution of this equation is of the form: 
\begin{equation}
P_L(m;t)=\sum_{m=0}^{\infty}b_mP_L^me^{\epsilon_mt}
\end{equation}
The eigenvalues, given by the coefficients of $P_L(m;t)$ are:
\begin{equation}
\epsilon_m=-[( n_L-m)+\frac{(\lambda+1)(z-n_L\lambda ^{L-1})}{1-\lambda(z-1)}]
\end{equation}

For the assumptions made, these eigenvalues are always negative, so the probability of having $m$ particles on shell $L$ while the other shells are empty decreases with time, as one might expect. Given initial conditions, and set system size,  we can find the corresponding eigenvectors  $P_L^m$.  To achieve the same goal,  we use a recursive method. For convenience, we use the following notations from now on:

\begin{eqnarray}
n_L&=&z(z-1)^{L-1}\\
\gamma&=& z \left( z-1 \right) ^{L-1}+{\frac {
 \left( \lambda+1 \right)  \left( z-z \left( z-1 \right) ^{L-1}{\lambda}^{L-1}
 \right) }{1-\lambda \left( z-1 \right) }}
\end{eqnarray}

Eq. 39 then becomes:

\begin{equation}
\frac{\partial P_L(m;t)}{\partial t}=(n_L-m+1)P_L(m-1;t)-(\gamma-m)P_L(m;t)
\end{equation}

First, we start with $m=0$:
\begin{equation}
{\frac {d}{dt}}{\it P_L} \left( 0;t \right) =- \gamma {\it P_L} \left(0; t \right)
\end{equation} 
We assume an empty initial configuration, so $P_L(0;0)=1$. The solution is:

\begin{equation}
P_L(0;t)={{\rm e}^{-\gamma\,t}}
\end{equation} 
With this solution, we go into the equation for $P_L(1;t)$ and solve it to get 
\begin{equation}
P_L\left(1;t \right) =n_{{L}} \left( 1-{{\rm e}^{-t}} \right) {
{\rm e}^{- \left( \gamma-1 \right) t}}
\end{equation} 

Following this method, we find a general form for the probability as:

\begin{equation}
P_{L}(m;t)={\frac { \Gamma  \left( m-{
\it n_L} \right)  \left( {{\rm e}^{-t}}-1 \right) ^{m}{{\rm e}^{-
 \left( \alpha-m \right) t}}}{\Gamma  \left( -{\it n_L} \right) m!
}}
\end{equation}

Fig. 8(b) displays the probability distribution of the deposition of two drug particles ($m=2$) on the final shell of a five generation ($L=5$) Cayley tree with $z=3$ and arbitrary inner-shell-dependent deposition rate of $r(l) = 0.4^{l}.$ Note that $r(L)=1$. The probabilities here are two orders of magnitude higher than those of the third distribution for the same m (see Fig. 8(a)).  The probability reaches a maximum of roughly $0.17$ at $t=0.037$, again in arbitrary units.\\

\bf b) Monomer deposition on interior shells $l<L$: variable rates \rm
\\

The same steps can be followed to solve for $P_l(m;t)$, the probability of having $m$ particles on shell $l<L$, provided that the rest of the shells are empty. In this case, the master equation that needs to be solved is:
\begin{eqnarray}
\frac{\partial P_l(m;t)}{\partial t}&=&r(l)z(n_l-m+1)P_l(m-1;t)-[n_L-mzr(l)\\
&+&\sum_{l'=1}^{L-1} n_{l'}(r(l')+r(l'-1))]P_l(m;t)\nonumber
\end{eqnarray}
 with $n_l$ being the number of sites on the $l^{th}$ shell.

The equation can be written in a more compact form:

\begin{eqnarray}
\frac{\partial P_l(m;t)}{\partial t}&=&\lambda^lz(n_l-m+1)P_l(m-1;t)-[n_L-mz\lambda^{l}\\
&+&\frac{(\lambda+1)(z-n_L\lambda ^{L-1})}{1-\lambda(z-1)}]P_l(m;t)\nonumber
\end{eqnarray}
The first term on the right hand side represents the process of adding a particle in the $l$ shell.  There are $(n_l+1-m)$ empty sites in the $l$ shell before the addition. Each such site is part of $z$ pairs, and the rate at which particles attached to $l$-shell sites is $r(l)$. Thus, the term has the form $+(n_l+1-m)z r(l)P_l(m-1)$.  Now, adding an $L$-shell particle when there are no particles there can happen in any of the $n_L$ empty sites, each of which is part of only one pair. The filling rate is $1$, producing the next term $-n_L P_l(m)$.   Each inner shell $l'$ (except shell $l$) is totally empty, thus has $n_{l'}$ fillable sites, each a part of $z$ pairs, filled at rate $r(l')$, producing $-n_{l'} z r(l') P_l(m)$.  For shell $l$ this rate is $-(n_{l'}-m) z r(l') P_l(m)$.

In terms of $\gamma$, the equation can be rewritten as:

\begin{equation} 
\frac{\partial P_l(m)}{\partial t} = (n_l+1 - m) z \lambda^{l} P_l(m -1)-(\gamma - m z \lambda^{l} )P_l(m ) . 
\end{equation}

The matrix associated with the master equation is tridiagonal, and the eigenvalues are the coefficients of $P_l(m;t)$:
\begin{equation}
\epsilon_m=-(\gamma - m z \lambda^{l} )
\end{equation}

Using the same method as in the previous section, we find for $P_l(m;t)$ the following solution:

\begin{equation}
P_l(m;t)={\frac {\Gamma  \left( m-{\it n_l} \right)  \left({{\rm e}^{-t{\lambda}^{
l}z}} -1\right) ^{m}{{\rm e}^{- \left( a-zm{\lambda}^{l} \right) t}}}{\Gamma 
 \left( -{\it n_l} \right) m!}}
\end{equation}
 with $n_l=z(z-1)^{l-1}$.

Fig. 8(a) displays the probability of deposition of two drug particles ($m=2$) on the third shell of a five generation ($l=3$, $L=5$) Cayley tree with $z=3$, and arbitrary inner-shell-dependent deposition rate $r(l)=0.4^{l}$. The probability reaches a maximum of roughly $3.6\times10^{-4}$ at $t=0.037$.

In general, as seen in the previous section for the deposition with constant rates,
 \begin{equation}
P(m_1;m_2;..m_{\ell};m_L;t)=\frac{\prod_{i\in\{1,2,..\ell\}} P(m_i;t)}{P(0;0;t)^{L_{total}}}
\end{equation}
where $\ell$ is the set of interior occupied shells (with at least one empty shell in between), and $L_{total}$ is the total number of occupied levels (including the last external shell).

\section{Summary and open questions}
In this paper we presented a class of adsorption models on Cayley trees with constant and variable attachment rates, and their possible applications for ionic self-assembly of thin films and drug encapsulation. Fig. 9 illustrates a sample comparison between the two cases (constant and variable rates) for the probability of having four particles on an internal shell $\ell=3$ for a generation five tree. First note that $r_3$ is set to zero so as to block the possibility of dimer deposition (as was considered in Section 3) for better comparison with the shell-dependent rate model. The shape of each graph is essentially identical. Each has a zero probability at the initial time (at $t=0$ the tree is empty), slopes up to a maximum, and then completes the trace of a bell skewed slightly to the right and approaching zero for large times. Graphs 9(a) and 9(b) differ only in their $r_2$ value. For Fig. 9(b), $r_2 = r_1 = 0.2$, meaning we model no electrostatic screening in the deposition rates, whereas for Fig. 9(a), $r_2 = 2r_1 = 0.4$, meaning that a deposition with no neighbor is twice as likely as one with neighbors.  The third shell of a Cayley tree with coordination number $z=3$ has twelve open nodes for the $m=4$ particles. One expects $r_2$ to have a very large impact on the distribution since there are three times as many spaces as particles. The maximum probability occurs at $t=0.0565$ for (a) and $t=1.11$ for (b), with a difference of roughly 49.1\%. Halving the $r_1$ value roughly doubles the time scale. The maximum probability remains almost unchanged, with a percent difference of about 7.1\%. The time scale is what is most affected, with the probability of the $m=4$ state for the third shell being more spread out for (b).

The $\lambda$ values in (c) and (d) are chosen to match the $r_2$ values in (a) and (b) respectively. This does not make the deposition rate in (c) or (d) equal to those of (a) or (b) because $r(l) = \lambda^{l}$, so the rates for the third shell are 0.064 and 0.008 respectively. These are much lower and explain the maximum probabilities, which are four and seven orders of magnitude lower than those for the constant rate graphs. The deposition rates would be comparable in the first shell, for $l=1$. For all graphs, we can estimate the maximum probability and the time at which it occurs. For variable rates, the peaks of the distributions are of the order of $10^{-7}$ (for $\lambda=0.4$) and, $10^{-11}$ (for $\lambda=0.2$), practically negligible. The same type of analysis can be done for more general particle deposition patterns, where multiple shells are filled with a set number of particles.

The minimalist cooperative sequential models presented here can be generalized to incorporate more realistic elements. In our future work we consider a cooperative sequential adsorption model with deposition rates dependent on the number of all occupied neighbors. We also want to answer the following basic question: ``Under what circumstances is a Cayley tree a good enough approximation for a fully two-dimensional lattice?". For the models presented, when $r_1 = r_3=0$ ( monomers are blocked to adsorb if a neighbor is occupied; no dimers allowed), the particle coverage of the jammed state is quite different for the tree (75\%) and the lattice (36\%). When the occupation of the neighboring sites doesn't matter, $r_1=r_2=1$, the analytical solution for the particle density as a function of time matches perfectly the Monte Carlo simulations for a two-dimensional lattice. 

For the ISAM problem, the size of the particles has to be addressed, as well as possible diffusion of particles on the substrate and evaporation. For the drug encapsulation problem, it would be interesting to model the release of the drug molecules from the dendrimers and their interactions with the targeted tissue, as well as the mobility of the dendrimer branches. These models can also be used to address other kinds of problems related to voting behavior and the spread of epidemics. For an epidemics model, for example, the "occupied" nodes can play the part of an infected individual, while the open nodes are represented by the susceptible individuals. This would be the case of a standard susceptible-infected (SI) model. The model can also incorporate a recovery rate, to become a susceptible-infected-recovered (SIR) model. A version of this model on a Cayley tree is presented in \cite{tome}.

\section{Acknowledgments}

Special thanks to H. T. Williams, D. Shuler, and J. Liu for helpful conversations. This research was supported in part by the National Science Foundation under grant no. 1126360.

\begin{figure}[tbp] 
  \centering
  \includegraphics[bb=11 13 601 780,width=5.67in,height=7.37in,keepaspectratio]{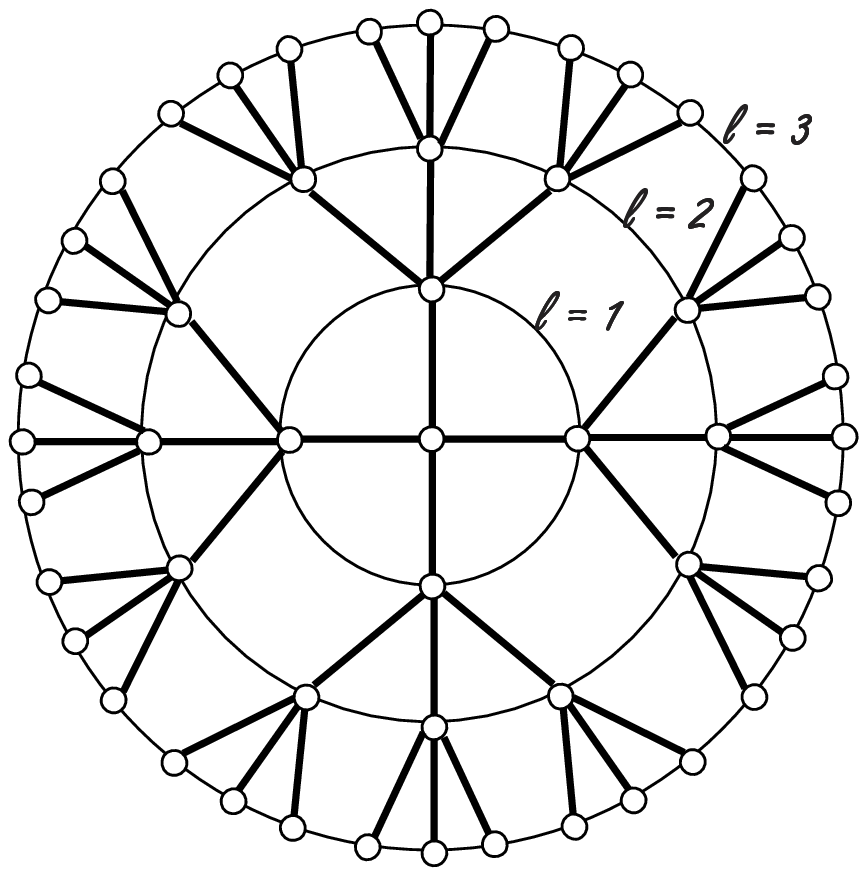}
  \caption{A Cayley tree of coordination number $z=4$ and generation $\ell=3$.}
  \label{fig:Fig1}
\end{figure}

\begin{figure}[bp] 
  \centering
  \includegraphics[bb=0 0 1024 768,width=10.67in,height=9.36in,keepaspectratio]{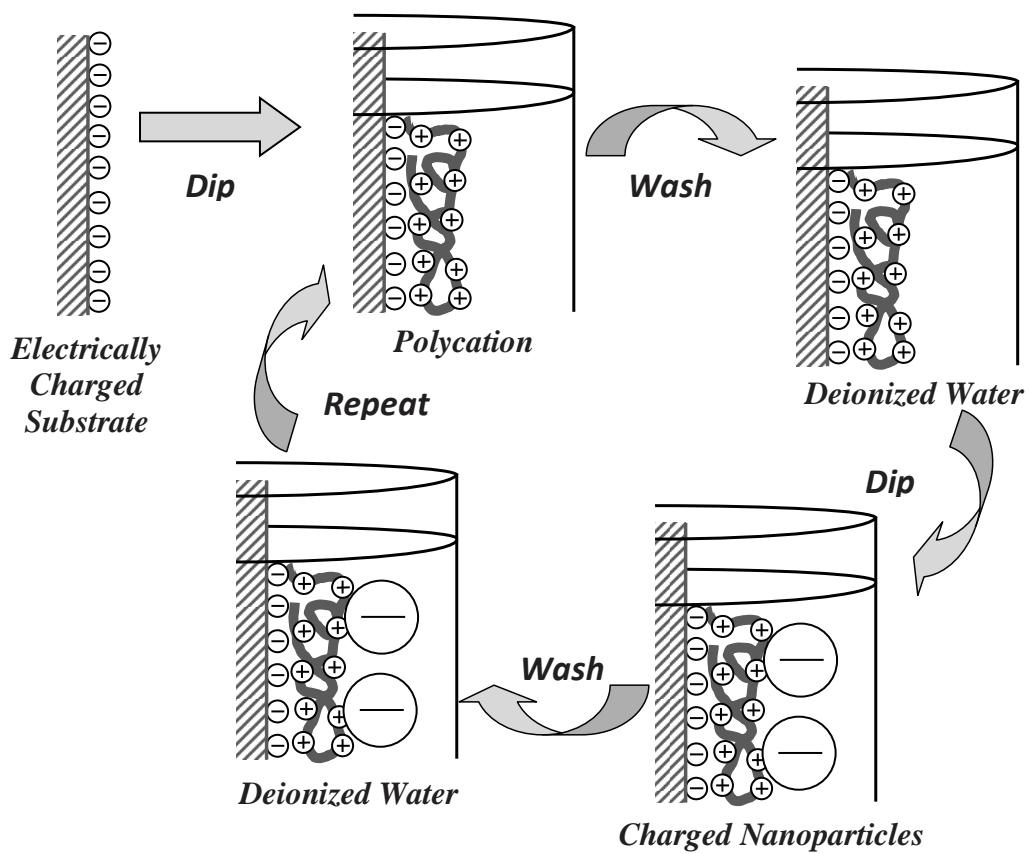}
  \caption{Basic steps followed in the layer-by-layer self-assembly process.}
  \label{fig:Fig2}
\end{figure}

\begin{figure}[tbp] 
  \centering
  \includegraphics[bb=0 0 1024 768,width=11.67in,height=8.37in,keepaspectratio]{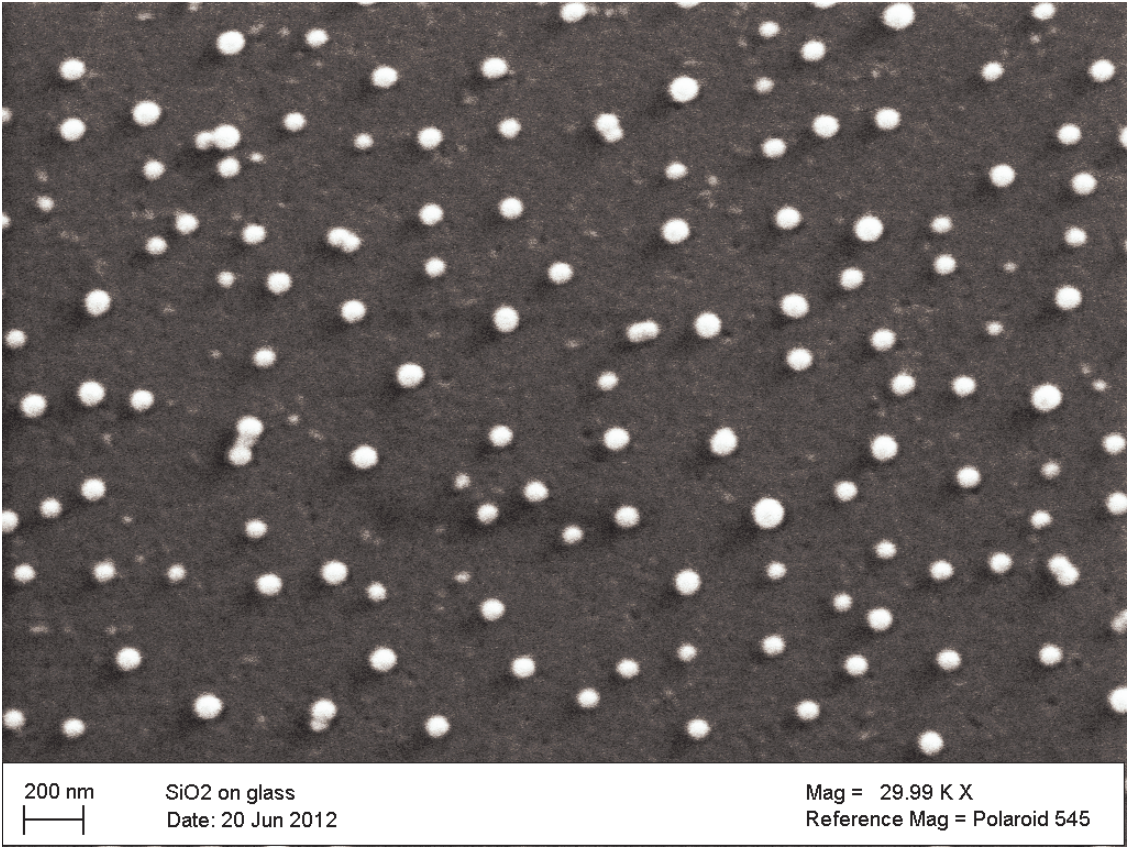}
  \caption{Sample SEM micrograph of a one-bilayer $SiO_2/PDDA$ coating made with silica particles of 45 nm average diameters.}
  \label{fig:Fig3}
\end{figure}

\begin{figure}[tbp] 
  \centering
  \includegraphics[bb=11 13 601 780,width=5.67in,height=7.37in,keepaspectratio]{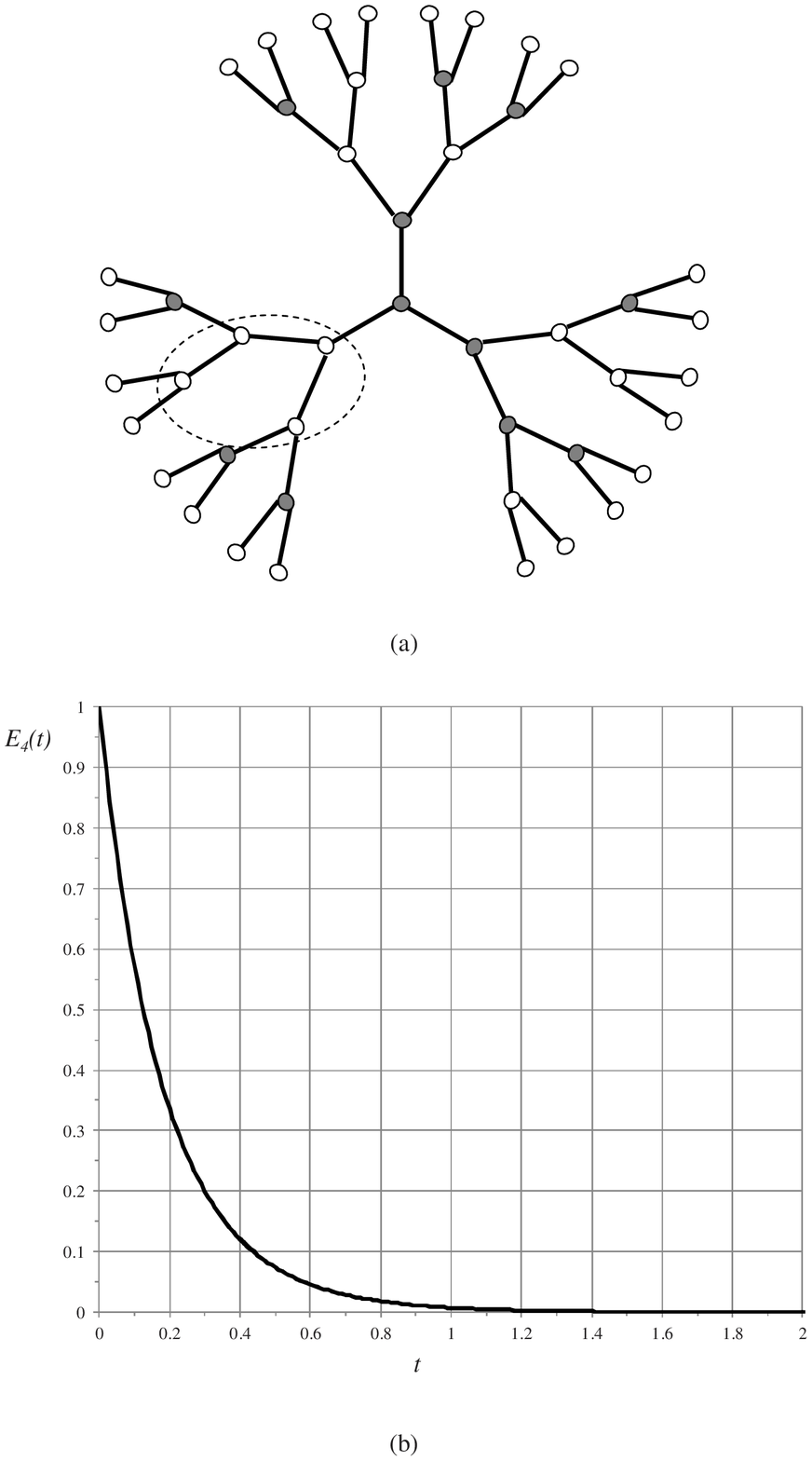}
  \caption{(a) Example of a cluster of four connected empty sites, $E_4$; (b) Time dependence of  the probability of having a cluster of four connected empty sites for a Cayley tree with $z=3$ and arbitrary deposition rates $r_1=0.2$, $r_2=0.4$ and $r_3=0.1$. }
  \label{fig:Fig4}
\end{figure}

\begin{figure}[tbp] 
  \centering
  \includegraphics[bb=11 13 601 780,width=5.67in,height=7.37in,keepaspectratio]{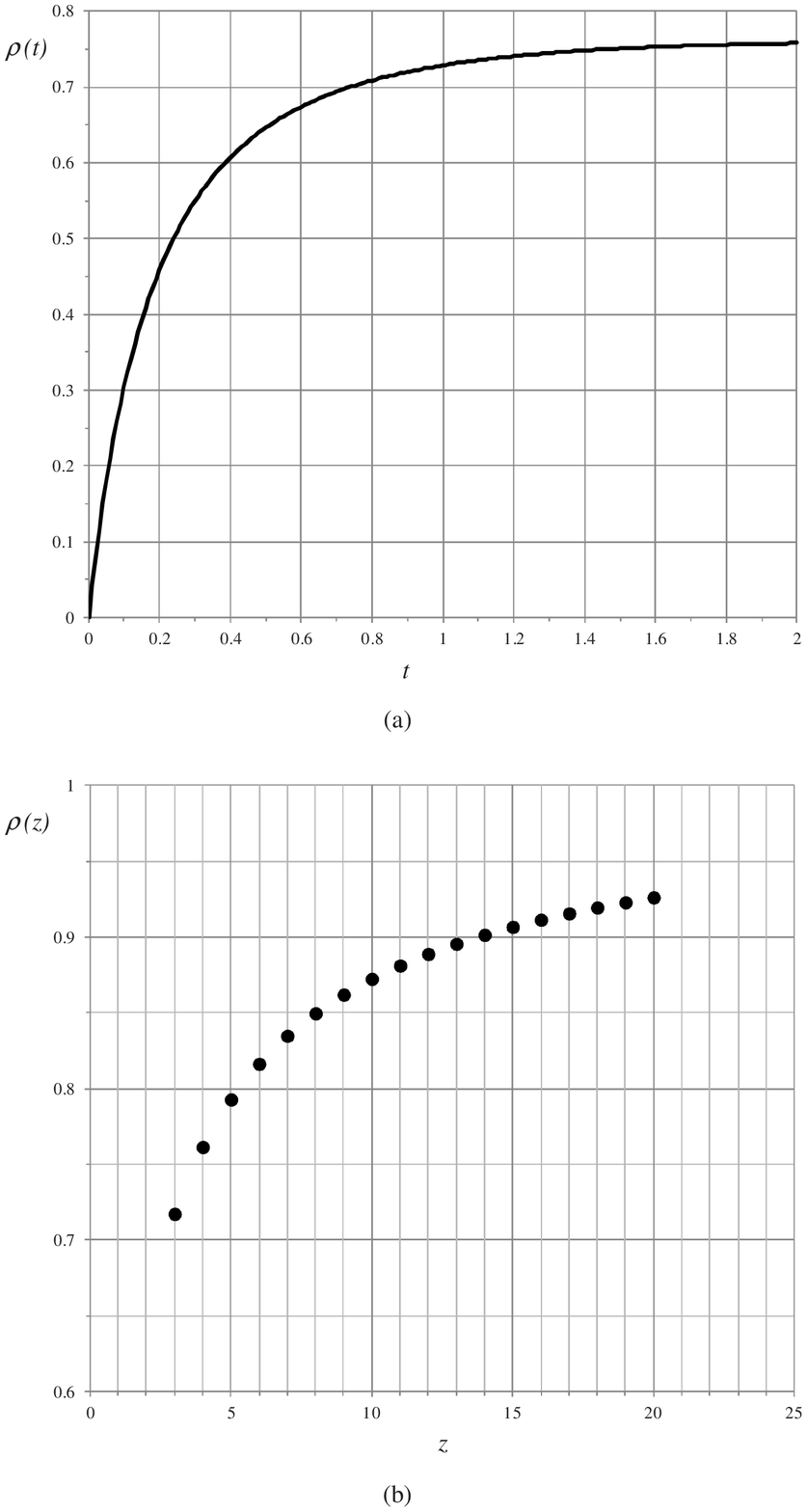}
  \caption{(a) Time dependence of the particle density for a Cayley tree with $z=4$, $r_1=0$, $r_2=1$, and $r_3=0.1$; (b) Steady state coverage for discrete values of coordination number z.}
  \label{fig:Fig5}
\end{figure}

\begin{figure}[tbp] 
  \centering
  \includegraphics[bb=11 13 601 780,width=5.67in,height=7.37in,keepaspectratio]{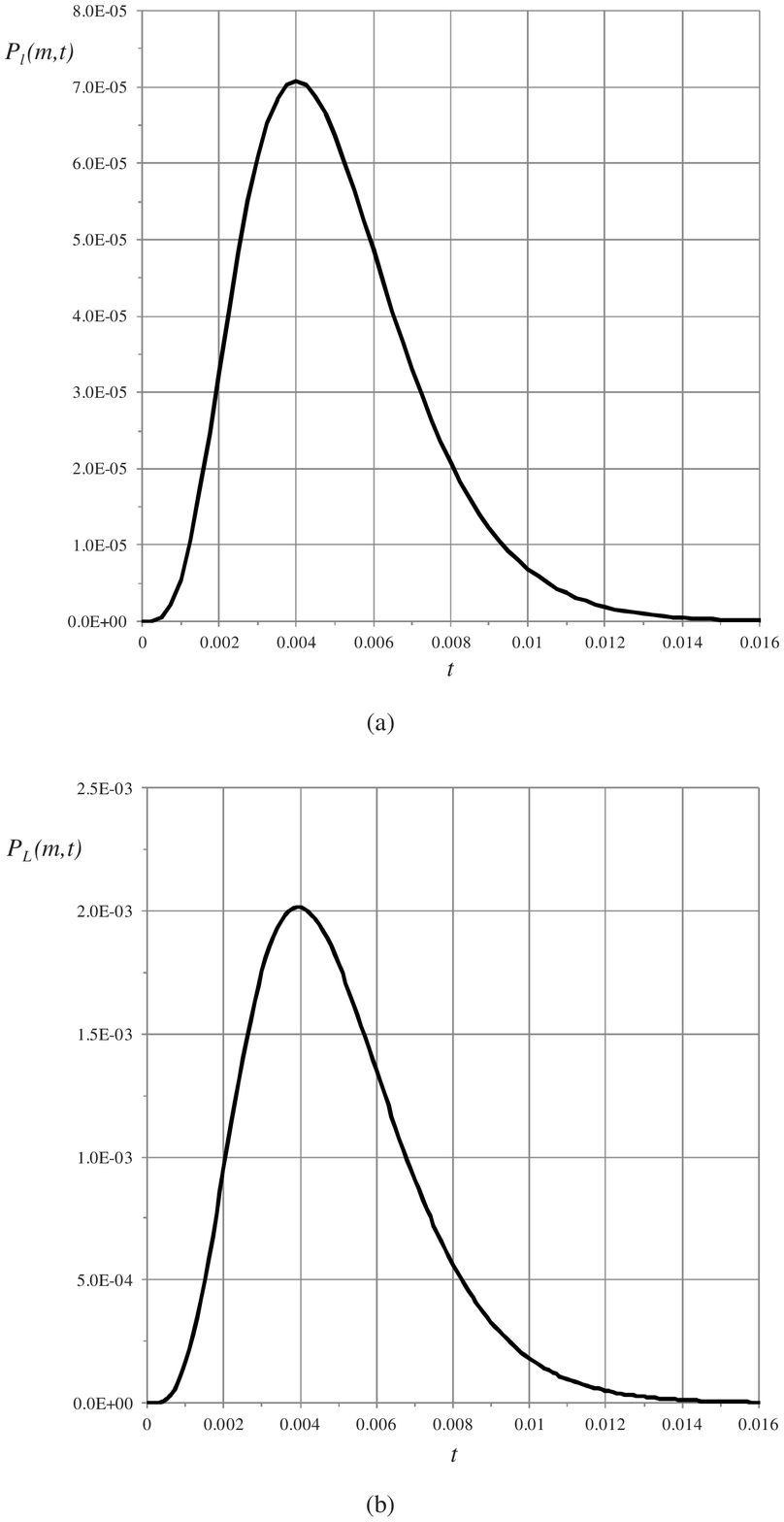}
  \caption{(a) Time dependence of the probability of having $m=4$ particles on the inner shell $\ell=3$ of a generation-five Cayley tree, with all the other shells empty, for $z=4$, $r_1=0$, $r_2=1$ and $r_3=0.1$; (b) Time dependence of the probability of having $m=4$ particles on the outer shell $L=5$ of a generation-five Cayley tree, with all the other shells empty, for $z=4$, $r_1=0$, $r_2=1$, and $r_3=0.1$. }
  \label{fig:Fig6}
\end{figure}

\begin{figure}[tbp] 
  \centering
  \includegraphics[bb=11 13 601 780,width=5.67in,height=7.37in,keepaspectratio]{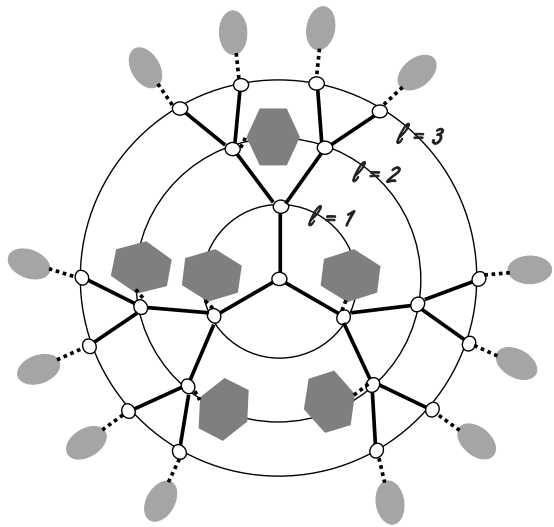}
  \caption{Sketch of the drug encapsulation and drug attachment on the outer shell of a generation-three dendrimer; different shaped enclosed particles signify the possibility of the dendrimer to carry different types of drug molecules. }
  \label{fig:Fig7}
\end{figure}

\begin{figure}[tbp] 
  \centering
  \includegraphics[bb=11 13 601 780,width=5.67in,height=7.37in,keepaspectratio]{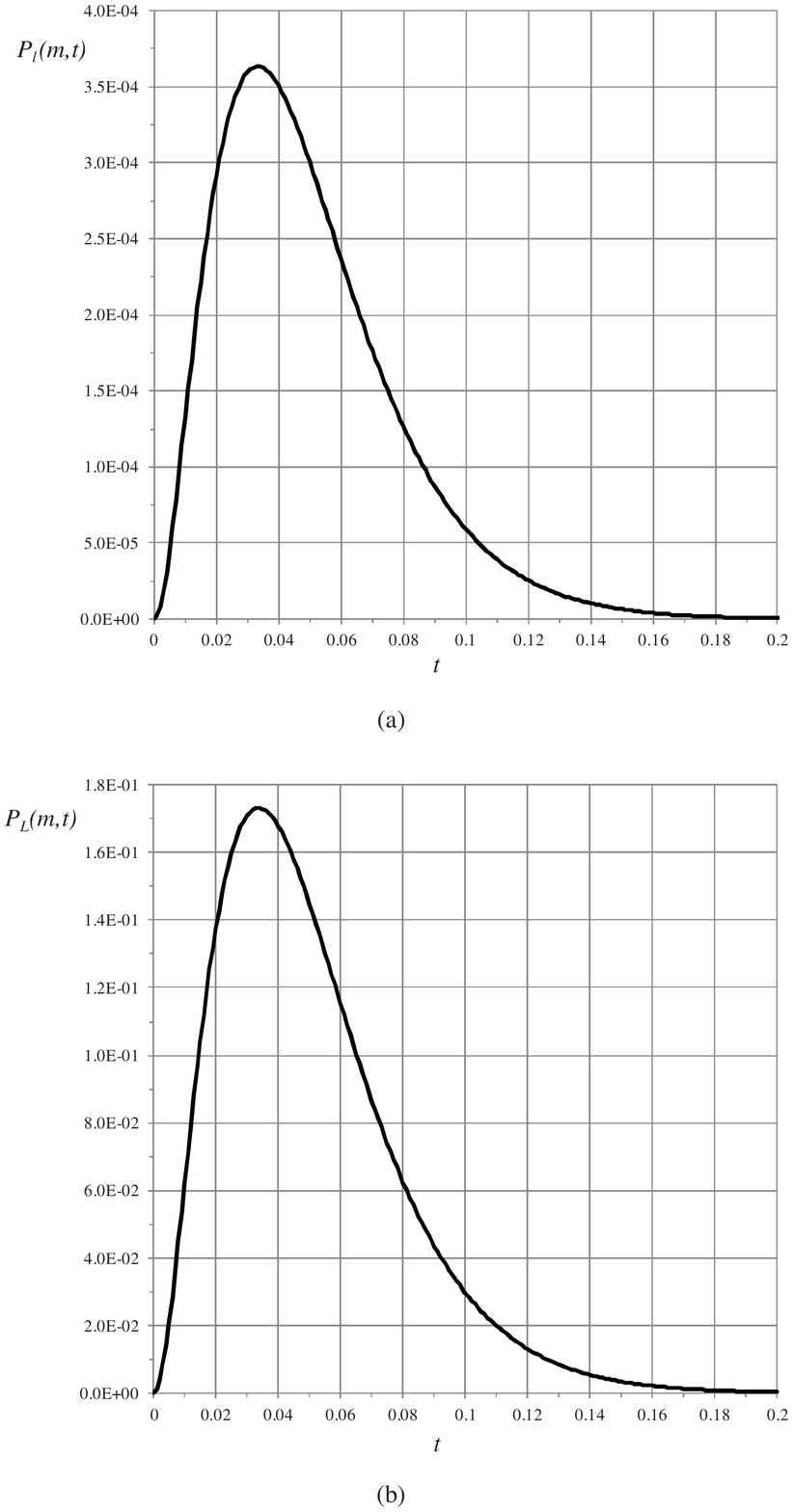}
  \caption{(a) Time dependence of the probability of having $m=2$ particles on the inner shell $\ell=3$ of a generation-five Cayley tree, with all the other shells empty, for $z=3$ and $\lambda=0.4$; (b) Time dependence of the probability of having $m=2$ particles on the outer shell $L=5$ of a generation-five Cayley tree, with all the other shells empty, for $z=3$ and $\lambda=0.4$. }
  \label{fig:Fig8}
\end{figure}

\begin{figure}[tbp] 
  \centering
  \includegraphics[bb=11 13 601 780,width=5.67in,height=7.37in,keepaspectratio]{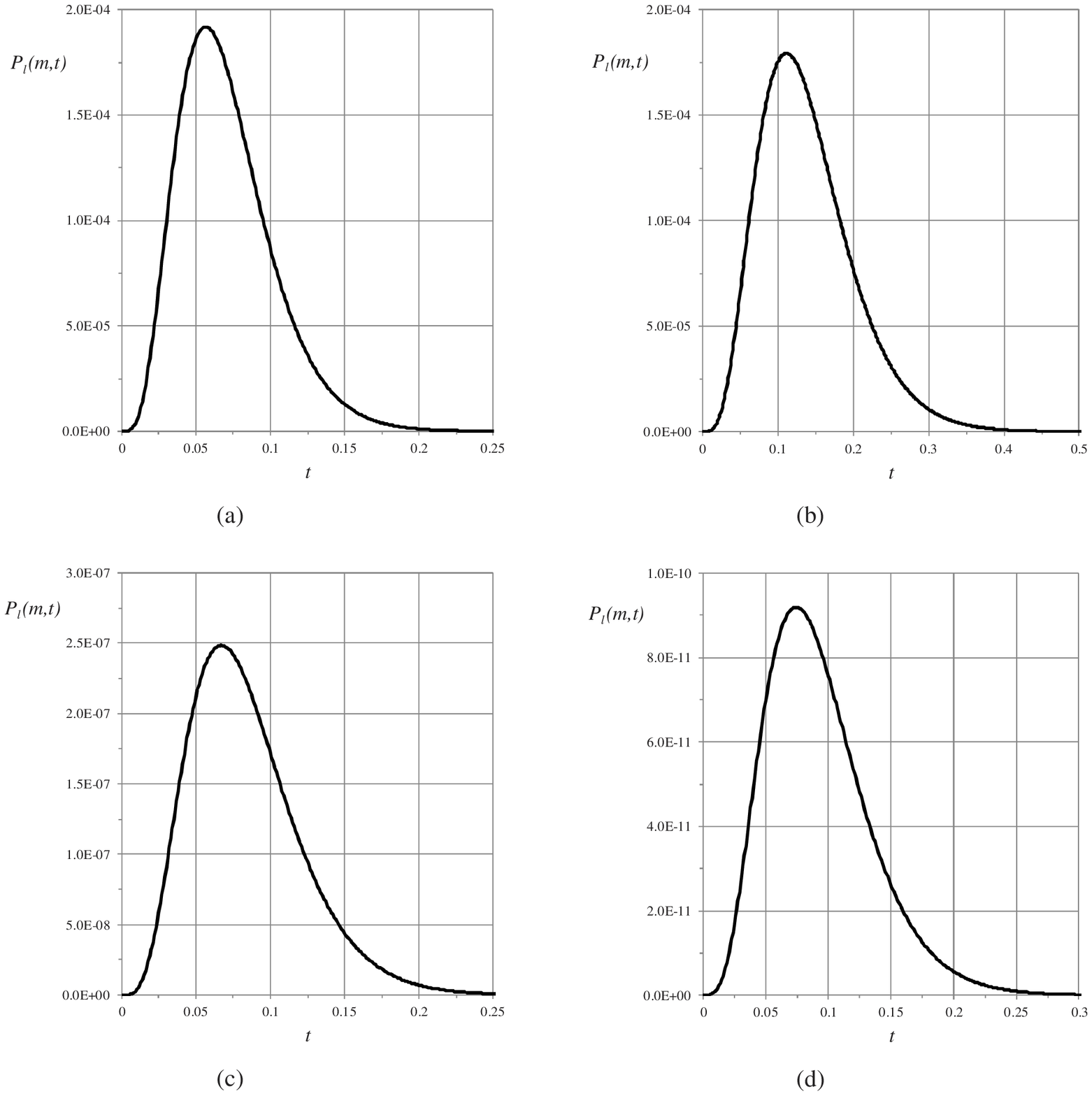}
  \caption{Sample comparison of time-dependent probabilities of having $m=4$ particles on the internal shell $\ell=3$ for a generation-five tree with $z=3$ for constant rates (a), (b) and variable rates (c), (d).}
  \label{fig:Fig9}
\end{figure}


\section{References}
\begin{thebibliography}{99}

\bibitem{heflin} Di Ventra M, Evoy S, and Heflin J R, \textit{Introduction to nanoscience and technology}, New York: Springer (2004.) 
\bibitem{evans} Evans J W, \textit{Rev. Mod. Phys.}, \textbf{65}, 1281 (1993.)
\bibitem{privman} Privman V, \textit{Nonequilibrium statistical mechanics in one dimension},  Cambridge
University Press, Cambridge, England  (1997.)
\bibitem{cadilhe1} Cadilhe A and Privman V, \textit{Mod. Phys. Lett. B} \textbf{18}, 207 (2004.)
\bibitem{cadilhe} Cadilhe A, Araujo N A M, and Privman V, \textit{J. Phys. Cond. Mat.} \textbf{19} (2007.)
\bibitem{matin} Matin L F, Aghamohammadi A, and Khorrami M, \textit{Eur. Phys. J. B}, \textbf{56}, 243 (2007.)
\bibitem{ali} Alimohammadi A and Olanj N, \textit{Physica A}, \textbf{389}, 549 (2010.)
\bibitem{gouet} Gouet R and Sudbury A, \textit{J. Stat. Phys.}, \textbf{130}, 935 (2008.)
\bibitem{iler} Iler R K, \textit{J. Colloid Interface Sci.}, \textbf{21}, 569 (1966.)
\bibitem{ritter} Yancey S E, Zhong W, Heflin J R, and Ritter A L, \textit{J. Appl. Phys.} \textbf{99}, 034313 (2006.)
\bibitem{dendri1} Hecht S and Frechet J M J, \textit{Angew. Chem. Int. Ed.} \textbf{40}, 74 (2001.)
\bibitem{dendri2}Kolhe P, Misra E, Kannana R M, Kannanb S, and Lieh-Lai M, \textit{I. J. Pharm.} \textbf{259}, 143 (2003.)
\bibitem{ostilli} Ostilli M, \textit{Physica A} \textbf{391}, 3417 (2012.)
\bibitem{isam1} Decher G, Hong J D, and Schimitt J, \textit{Thin Solid Films} \textbf{210}, 831 (1992.)
\bibitem{isam2}Fou A C, Onitsuka O, Ferreira M, Rubner M F, and Hsieh B R, \textit{J. Appl. Phys.},\textbf{79}, 7501 (1996); Onitsuka O, Fou A C, Ferreira M,  Rubner M F, and Hsieh B R, \textit{J. Appl. Phys.}, \textbf{80}, 4067 (1996.)
\bibitem{ben avraham} ben-Avraham D and Glasser M L, \textit{J. Phys. Cond. Mat.} \textbf{19} (2007.) 
\bibitem{redner} Krapivsky P L, Redner S, and Ben-Naim E V, \textit{A kinetic view of statistical physics},  Cambridge
University Press, Cambridge, England  (2010.)
\bibitem{tome}  Tome T and Oliveira M J, \textit{J. Phys. A: Math. Theor.} \textbf{44}, 095005 (2011.)
\end{thebibliography}
\end{document}